\documentclass[runningheads]{llncs}
\usepackage{graphicx}
\usepackage{tikz}
\usepackage{tkz-graph}
\usepackage{bbm}
\usepackage{booktabs}
\usepackage{listings}
\usepackage{amsfonts}
\usepackage{amsmath}
\usepackage{wrapfig}
\usepackage{pgfplots}
\usepackage{pgfplotstable}
\usepackage{adjustbox}

\newtheorem{mydef}{Definition}

\begin{document}
\title{Extracting Multiple Viewpoint Models from Relational Databases}
\titlerunning{Extracting MVP Models from Databases}
%
%
\author{Alessandro Berti\inst{1}\orcidID{0000-0003-1830-4013} \and
Wil van der Aalst\inst{1}\orcidID{0000-0002-0955-6940}}
%
%
\institute{Process and Data Science department, Lehrstuhl fur Informatik 9 52074 Aachen, RWTH Aachen University, Germany}
\maketitle              
\begin{abstract}
Much time in process mining projects is spent on finding and understanding data sources and extracting the event data needed. 
As a result, only a fraction of time is spent actually applying techniques to discover, control and predict the business process.
Moreover, current process mining techniques assume a \emph{single} case notion. 
However, in real-life processes often different case notions are intertwined. 
For example, events of the same order handling process may refer to customers, orders, order lines, deliveries, and payments.
Therefore, we propose to use \emph{Multiple Viewpoint (MVP) models} that relate events through objects and that relate activities through classes.
The required event data are much closer to existing relational databases. MVP models provide a holistic view on the process, 
but also allow for the extraction of classical event logs using different viewpoints. 
This way existing process mining techniques can be used for each viewpoint without the need for new data extractions and transformations.
We provide a toolchain allowing for the discovery of MVP models (annotated with performance and frequency information) from relational databases.
Moreover, we demonstrate that classical process mining techniques can be applied to any selected viewpoint.
\end{abstract}
\section{Introduction}
\label{sec:introduction}

Process mining is a growing branch of data science that aims to extract insights from event data recorded in information systems. 
Examples of process mining techniques include
process discovery algorithms that are able to find descriptive process models, 
conformance checking algorithms that compare event data with a given process model to find deviations, 
and predictive algorithms that use the discovered process model to anticipate bottlenecks or compliance problems. 
Gathering high-quality event data is a prerequisite for the successful application of process mining projects.
However, event data are often hidden in existing information systems (e.g., the ERP systems of SAP, Microsoft, Oracle). 
Most systems are built on top of relational databases, to ensure data integrity and normalization. 
Relational databases contain entities (tables) and relations between entities.
Events correspond to updates of the database, i.e., changes of the ``state'' of the information system.
These updates may have been stored in tables of the databases (through {\it in-table versioning} like the change tables in SAP or any table that contains dates or timestamps) or
can be retrieved using some database log (like {\it redo logs} \cite{yano2013practical,de2017redo} explicitly storing all database updates).

Extracting events from a database requires domain knowledge and may be very time-consuming. 
One of the reasons is that process mining techniques require a classical event log where each event needs to refer to a case.
The case notion is used to ``correlate'' events and the corresponding process model shows the life-cycle model for the selected case notion.
However, in the same process there may be suppliers, customers, orders, order lines, deliveries, and payments. 
These correspond to objects of the class model describing the database. One event may refer to a subset of such objects. A payment may refer to the payment itself, a customer, and an order.
A delivery may refer to the delivery itself, but also to a subset of order lines or even multiple orders.
Several views on the database could be retrieved, this means that, for the same database, several event logs and process models need to be extracted.
When the process involves many different entities, the construction of the view is not easy.
While there are some methods to automatically infer a/some case notion(s) from unstructured data \cite{bayomie2016deducing,helal2015runtime,burattin2011framework},
in most cases the specification happens manually.
Moreover, the data extractions and transformations may be time-consuming and one quickly loses the overview. This often leads to divergence and convergence errors: 
events are forgotten or inadvertently duplicated leading to incorrect conclusions.
\begin{figure*}
\centering
\includegraphics[width=0.4\paperheight]{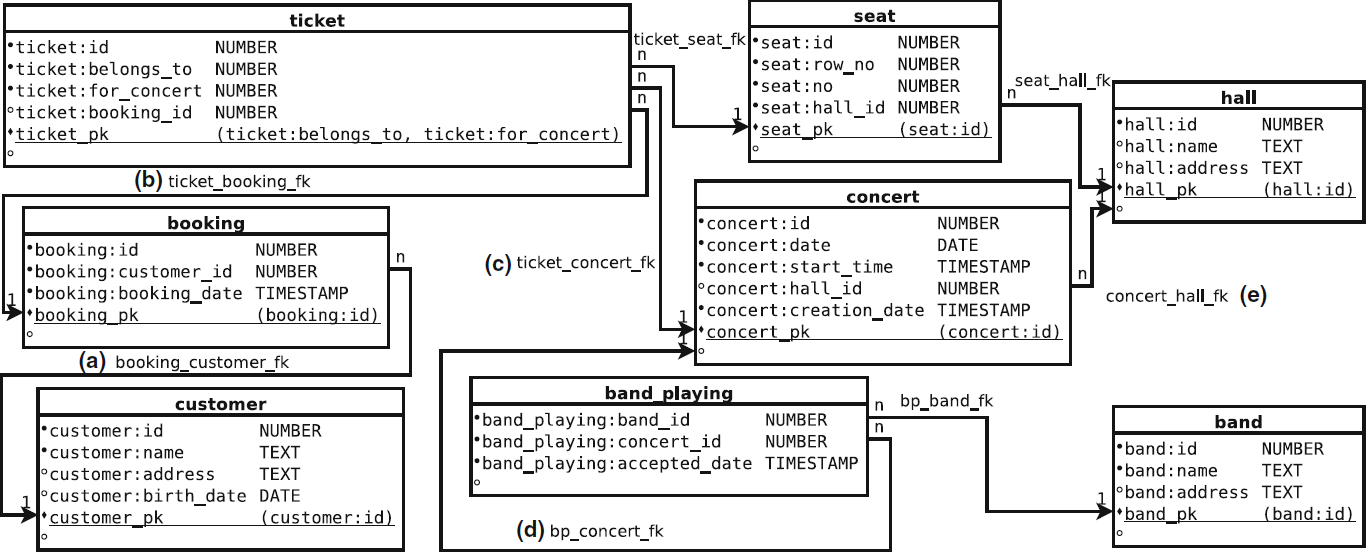} \\
(a) Database model of a concert database.\\
\vspace{10px}
\includegraphics[width=0.5\paperheight]{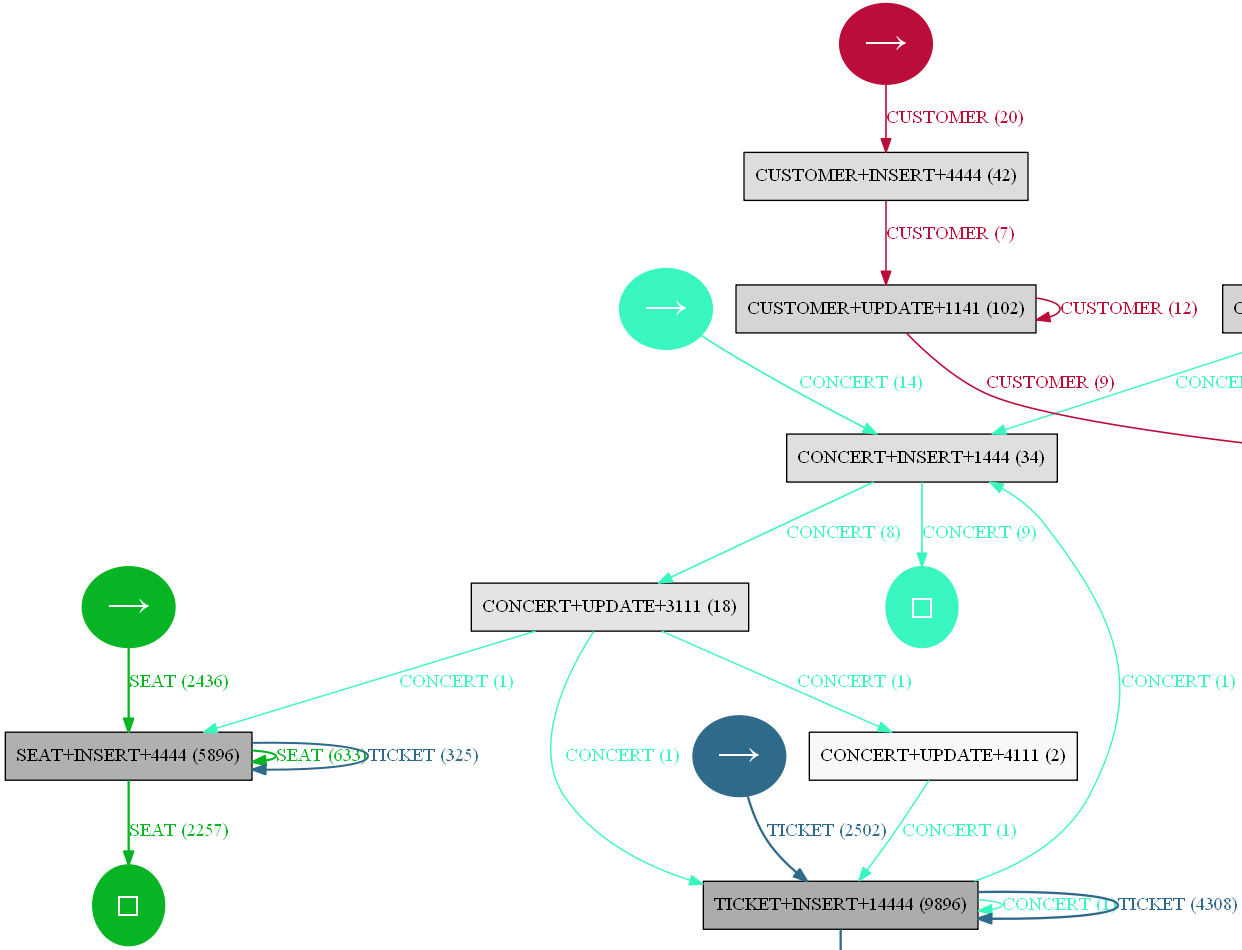} \\
(b) MVP model annotated with frequencies.\\
\vspace{10px}
\caption{Illustration of the approach using an example database taken from \cite{de2018connecting}.
The classes are highlighted using different colors and labels on arcs. Based on a viewpoint (i.e., a set of classes) a classical event log can be generated and analyzed.}\label{fig:concertFrequency}
\end{figure*}



This paper introduces a new modeling technique that is able to calculate a graph where relationships between activities are shown without forcing the user to specify a case notion,
since different case notions are combined in one succint diagram.
The resulting models are called MVP models. Such models belong to the class of artifact-centric models \cite{cohn2009business,nigam2003business} that combine
data and process in a holistic manner to discover patterns and check compliance \cite{lohmann2011compliance}.
MVP models are annotated with frequency and performance information (e.g., delays) and provide a holistic view on the whole process.
The colors of the relationships refer to the original classes. Using frequency based filtering and selections the MVP model can be seamlessly simplified.
Any non-empty subset of classes provides a \emph{viewpoint}. Given a viewpoint and an MVP model, we can automatically generate a classical event log and apply existing process mining techniques.
Hence, the holistic view of the MVP model is complemented by detailed viewpoint models using conventional notations like Petri nets, BPMN models, process trees, or simple directly-follows graphs.

The techniques have been implemented using PM4Py (\url{pm4py.org}). In-memory computation is used to handle large data sets quickly.
The approach has been evaluated using a log extracted from a real-life information system, and has proven to scale linearly with the number of events, classes and asymptotically linearly with the number of objects per class, while
the execution time grows quadratically with the number of activities. This is a stark contrast with existing techniques (like OCBC models) with significantly worse complexity.
 
In Section \ref{sec:selfAssessment}, an assessment on real-life database event logs is done.
Moreover, a comparison with some existing techniques (OpenSLEX \cite{de2018connecting,de2019process} and OCBC models \cite{van2017object}) is performed,
considering the execution time and the usability of these approaches.

The remainder of the paper is organized as follows. 
Section~\ref{sec:rw} presents related work.
In Section~\ref{sec:background} classical and database event logs are introduced.
Section~\ref{sec:approach} presents our approach to discover Multiple Viewpoint (MVP) models including Event-to-Object (E2O) graphs, Event-to-Event (E2E) multigraphs, and Activity-to-Activity (A2A) multigraphs.
Section~\ref{sec:projection} introduces the notion of viewpoints and the automatic creation of classical event logs.
Section \ref{sec:tool} presents our implementation using PM4Py.
In Section \ref{sec:assessment}, we evaluate MVP models and our implementation by comparing results for real-life data sets with competing techniques.

\section{Related Work}
\label{sec:rw}


\begin{figure}[ht]
\centering
\includegraphics[width=340px]{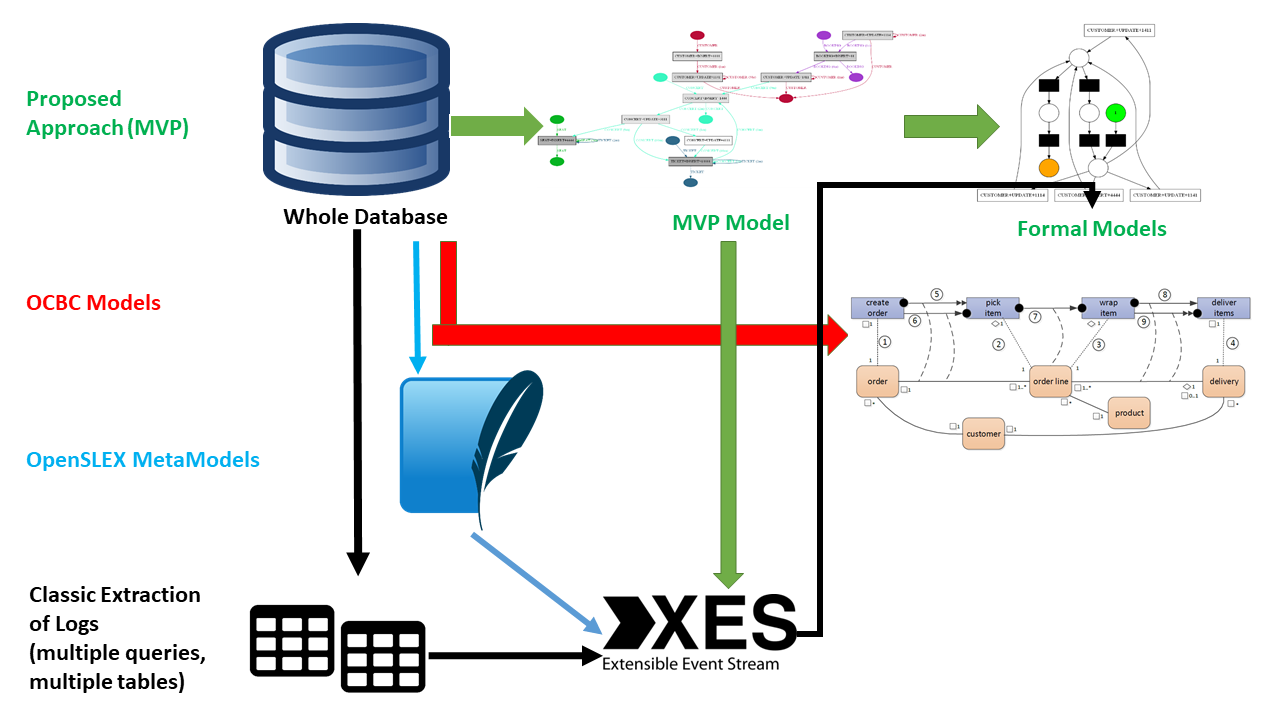}
\caption{Comparison of four different process mining ETL scenarios. Classical log extraction consider different queries, possibly targeting different tables,
to prepare event logs for classic process mining techniques.
OpenSLEX provides an easier access point to retrieve event data, but still requires the manual specification of database queries.
OCBC technique extracts a single process model from the database schema, that could be useful for the purpose of understanding and checking the schema, but provide no way to retrieve a classical event log. The MVP technique, on the other hand, provides an easy visualization of the relationships between activities on top of a database, moreover it provides the possibility to get formal models (directly from the MVP models) and event logs to use with classic process mining techniques.}
\label{fig:mvpIntroduction}
\end{figure}

Related work may be divided into different categories:
\begin{itemize}
\item Approaches to extract event data from databases and to make queries easier.
\item Representation of Artifact-centric models.
\item Discovery of process models combining several case notions.
\end{itemize}

A few example approaches are shown in Fig. \ref{fig:mvpIntroduction} and related to MVP models.

\subsection{Related Work: Extracting Event Data From Databases}

There has been earlier work on making SQL queries easier \cite{calvanese2017ontop,bouchou2014semantic}. The basic idea is to provide the business analyst a way to express queries in a simpler language (SPARQL query).
Some other papers related to database querying in the context of process management are \cite{momotko2004process,liu2009semantic,song2011querying,metzke2013enabling,tang2018querying}.

In \cite{nooijen2012automatic}, an automated technique to discover, for each notion of data object in the process, a separate process model that describes the evolution
of this object, is presented. The technique is based on relational databases, and decomposes the data source into multiple logs, each describing the cases of a
separate data object.

The OpenSLEX \cite{de2018connecting,de2019process} is an high-level meta-model, that permits easier queries, obtained from the raw database data scattered through tables (for example, the case identifier, the activity and the timestamp may be columns of different tables that need to be joined together).
The aim of OpenSLEX is to let user focus on the analysis, dealing only with elements such as events, activities, cases, logs, objects, objects versions, object classes and attributes that are introduced in \cite{de2018connecting}.
The meta-model could be seen as a schema that captures all the pieces of information necessary to apply process mining to database environments.
To obtain classical event logs, a case notion (connecting events to each other) needs to be used.
The OpenSLEX implementation provides indeed some connectors for database logs (redo logs, in-table versioning, or specific database formats \cite{de2019process,ingvaldsen2007preprocessing}).
In the implementation described in \cite{de2018connecting}, OpenSLEX is supported by an SQLite database.

\subsection{Related Work: Representation of Artifact-Centric Models}

Business artifacts (cf. \cite{cohn2009business,narendra2009towards}) combine data and process in an holistic manner as the basic building block. These correspond to key business entities which evolve as they pass through the business's operation.


In \cite{bhattacharya2007towards} a formal artifact-based business model, and declarative semantics based on the use of business rules, are introduced
along with a preliminary set of technical results on the static analysis of the semantics of an artifact-based business process model.

The Guard-Stage-Milestone (GSM) meta-model \cite{hull2010introducing,hull2011business}  is a formalism for designing business artifacts
in which the intended behavior is described in a declarative way, without requiring an explicit specification of the control flow.

Some approaches which focus on compliance checking are introduced in \cite{knuplesch2010enabling,ly2011monitoring}.
In \cite{knuplesch2010enabling}, support for data-aware compliance rules is proposed in a scalable way thanks to an abstraction approach that can serve
as preprocessing step to the actual compliance checking. In \cite{ly2011monitoring}, compliance rule graphs are introduced, that are a framework with support for root cause analysis,
and that can provide assistance in proactively enforcing compliance by deriving measures to render the rule activation satisfied.
In \cite{fahland2011behavioral}, conformance checking on artifact-centric processes is approached by partitioning the problem into behavioral conformance of single
artifacts and interaction conformance between artifacts, solving behavioral conformance by a reduction to existing techniques.

In \cite{steinau2018relational}, the concept of relational process structure is introduced, aiming to overcome some limitations of small processes such as artifacts,
object lifecycles, or proclets. In these paradigms, a business process arises from the interactions between small processes. However, many-to-many relationships
support is lacking. The relational process structure provides full support for many-to-many relationships and cardinality constraints at both design- and run-time.

\subsection{Related Work: Discovery of Process Models Using Several Case Notions}


\begin{figure}[ht]
\centering
\includegraphics[width=280px]{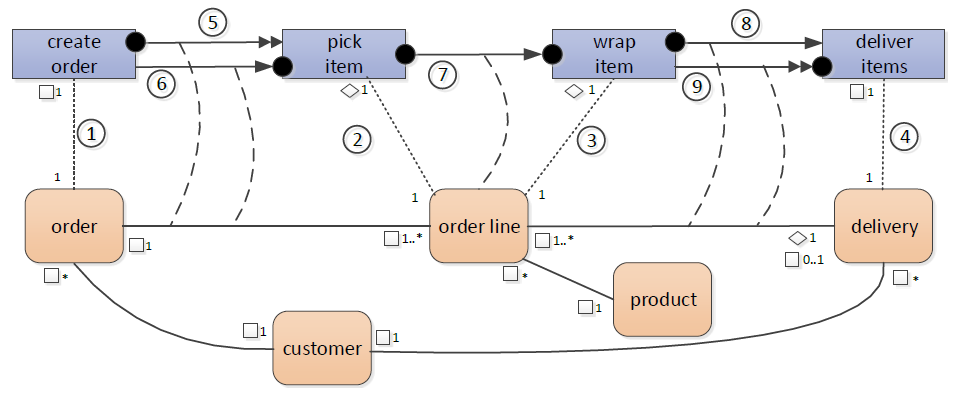}
\caption{Representation of a small Object-centric Behavioral Constraint (OCBC) model (taken from \cite{van2017object}).}
\label{fig:ocbcrep}
\end{figure}

In this category, we aim to describe some models that could be discovered by combining several case notions,
or studying the interactions between the different case notions:
interacting artifacts \cite{lu2015discovering},
multi-perspective models \cite{van2016discovering},
object-centric models \cite{van2017object}.

In \cite{lu2015discovering}, a semi-automatic approach is presented to discover the various objects supporting the system from the plain database of an ERP (Enterprise Resources Planning) system. An artifact-centric process model is identified describing the system's objects, their life-cycles, and some detailed
information about interactions between objects.

Multi-instance mining \cite{van2016discovering} was introduced to discover models where different perspectives of the process can be
identified. Instead of focusing on the events or activities that are executed in the context of a particular process, the focus is on the states
of the different perspectives and on the discovery of the relationships between them. The
{\it Composite State Machine Miner} \cite{van2016composite} supports the approach \cite{van2016discovering}. It quantifies and visualises the interactions between perspectives to provide additional process insights.

Object-centric models \cite{van2017object} are process models, involving entities and relations in the ER model, where multiple case notions may coexist.
A small OCBC model is represented in Fig. \ref{fig:ocbcrep}.
In the representation, the lower part (yellow) represents the class model, and the upper part represents the activities and the constraints between them. The activities and the
classes are connected by arcs if they are in relationship.
OCBC models can be discovered, and can be used for compliance checking, from XOC logs.
The XOC format is important because it extends the XES format with support to database-related information (related objects, relationships, state of the object model), and is one of the few choices to store
database event logs along with instances of the OpenSLEX meta-model.
XOC logs are using XML and contain a list of events. Each event is referring some objects, and contains
the status of the database at the moment the event happened.
In \cite{li2017automatic}, the algorithm to infer OCBC models is described,
that takes an XOC log as well as a set of possible behavioral constraint types as input, that means users can specify the constraint type set based on their needs.
ProM 6 plug-ins have been realized to import XOC logs, for the discovery of a process model, and for conformance checking on top of OCBC models.
The discovery algorithm can discover constraints of 9 different types.

\section{Database Event Logs}
\label{sec:background}

Relational databases are organized in entities (classes of {\it objects} sharing some properties), relationships (connections between entities \cite{chen1976entity}), attributes (properties of entities and relationships).
Events can be viewed as updates of a database (e.g. insertion of new objects, changes to existing objects, removal of existing objects).
Some ways to retrieve events from databases are:
\begin{itemize}
\item Using redo logs (see \cite{de2018connecting}). These are logs where each operation in the database is saved with a timestamp; this helps to guarantee consistency, and possibility to
rollback and recovery.
\item Using in-table versioning. In this case, the primary key is extended with a timestamp column. For each phase of the lifecycle of an object, a new entry is added to the in-table
versioning, sharing the base primary key values but with different values for the timestamp column.
\end{itemize}
An event may be linked to several objects (for example, the event that starts a marketing campaign in a CRM system may be linked to several customers), and an object may be linked to several events
(for example, each customer can be related to all the tickets it opens).
For the following definition, let $\mathcal{U}_E$ be the universe of events (all the events happening in a database context),
$\mathcal{U}_C$ be the universe of case identifiers,
$\mathcal{U}_A$ be the universe of activities (names referring to a particular step of a process),
$\mathcal{U}_{\textrm{attr}}$ be the universe of attribute names (all the names of the attributes that can be related to an event),
$\mathcal{U}_{\textrm{val}}$ be the universe of attribute values (all the possible values for attributes).


In this paper we consider event data closer to real-life information systems. Before providing a definition for database event logs, we define the classical event log concept.
\begin{mydef}[Classical Event Log]
\label{def:classicalEventLogDefinition}
A log is a tuple $L = (C_I, E, A, \textrm{case\_ev}, \allowbreak \textrm{act}, \allowbreak \textrm{attr}, \leq)$ where:
\begin{itemize}
\item $C_I \subseteq \mathcal{U}_C$ is a set of case identifiers.
\item $E \subseteq \mathcal{U}_E$ is a set of events.
\item $A \subseteq \mathcal{U}_A$ is the set of activities.
\item $\textrm{case\_ev} \in C_I \rightarrow \mathcal{P}(E) \setminus \{ \emptyset \}$ maps case identifiers onto set of events (belonging to the case).
\item $\textrm{act} \in E \rightarrow \mathcal{U}_A$ maps events onto activities.
\item $\textrm{attr} \in E \rightarrow (\mathcal{U}_{attr} \not\rightarrow \mathcal{U}_{val})$ maps events onto a partial function assigning values to some attributes.
\item $\leq ~ \subseteq E \times E$ defines a total order on events.
\end{itemize}
\end{mydef}
This classical event log notion matches the XES storage format \cite{verbeek2010xes}, that is the common source of information
for process mining tools like Disco, ProcessGold, Celonis, QPR, Minit, \ldots
An example attribute of an event $e$ is the timestamp $\textrm{attr}(e)(time)$ which refer to the time the event happened.
While, in general, an event belongs to a single case, in Def. \ref{def:classicalEventLogDefinition}
the function case\_ev might be such that cases share events.

For events extracted from a database,
the function case\_ev is not given, since an event may be related to different objects, and different case notions may exist.
In the following Def. \ref{def:newEventLogDefinition},
database event logs are introduced.

\begin{mydef}[Database Event Log]
\label{def:newEventLogDefinition}
Let $\mathcal{U}_O$ be the universe of objects (all the objects that are instantiated in the database context) and $\mathcal{U}_{OC}$ be the universe of object classes (a class defines the structure and the behavior of a set of objects).
A database event log is a tuple $L_D = (E, O, C, A, \textrm{class}, \textrm{act}, \textrm{attr}, \allowbreak \textrm{EO}, \leq)$ where:
\begin{itemize}
\item $E \subseteq \mathcal{U}_E$ is the set of events.
\item $O \subseteq \mathcal{U}_O$ is the set of objects.
\item $C \subseteq \mathcal{U}_{OC}$ is the set of object classes.
\item $A \subseteq \mathcal{U}_A$ is the set of activities.
\item $\textrm{class} : O \rightarrow C$ is a function that associates each object to the corresponding object class.
\item $\textrm{act} \in E \rightarrow A$ maps events onto activities.
\item $\textrm{attr} \in E \rightarrow (\mathcal{U}_{attr} \not\rightarrow \mathcal{U}_{val})$ maps events onto a partial function assigning values to some attributes.
\item $\textrm{EO} \subseteq E \times O$ relates events to sets of object references.
\item $\leq ~ \subseteq E \times E$ defines a total order on events.
\end{itemize}
\end{mydef}

This definition differs from Def. \ref{def:classicalEventLogDefinition}
because a case notion is missing (no \emph{case\_ev} function) and events are related to objects in a possible many-to-many relation.
A function \emph{EO} is introduced that relates events to sets of object references.
Moreover, the sets of objects and classes in the event log are specified, and a function \emph{class} that associates each object
to its class is introduced.

\section{Approach to Obtain MVP models}
\label{sec:approach}
In this section, the different ingredients of MVP models will be introduced. The E2O graph will be obtained directly from the database logs; the E2E multigraph will be
obtained in linear complexity by calculating directly-follows relationships between events in the perspective of some object; the A2A multigraph will be obtained in linear complexity
by calculating directly-follows relationships between activities in the perspective of some object class, using the information stored in the E2E multigraph. The A2A multigraph
is the main contributor to the visualization of the MVP model.
A projection function will be given in Section \ref{sec:projection} to obtain a classical event log when a so-called viewpoint is chosen.

\begin{figure}[ht]
\centering
\includegraphics[width=\textwidth]{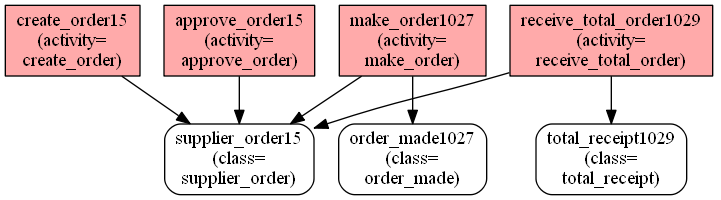}
\caption{Visualization of part of the E2O graph of an example database event log (found in the erp.xoc test file). Events (red nodes) are connected to objects (white nodes).}
\label{fig:e2ographvis}
\end{figure}
\begin{figure}[ht]
\centering
\includegraphics[width=\textwidth]{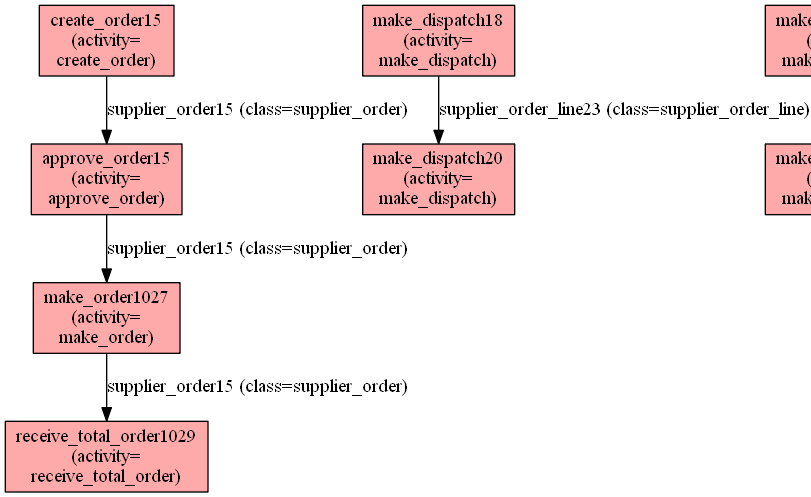}
\caption{Visualization of part of the E2E multigraph of an example database event log (erp.xoc test file). Events are connected to events; in the edge label, the (object) perspective has been reported.}
\label{fig:e2egraphvis}
\end{figure}

\label{sec:modelConstruction}
MVP models are composed of several graphs (E2O, E2E, A2A) and auxiliary functions (a complete definition will be presented at the end of this section), and are constructed by reading a representation of event data retrieved from a database (importing from intermediate structures like OpenSLEX or XOC logs).
\begin{mydef}[E2O Graph]
Let $L_D = (E, O, C, A, \allowbreak \textrm{class}, \allowbreak \textrm{act}, \allowbreak \textrm{attr}, \allowbreak \textrm{EO}, \allowbreak \leq)$ be a database event log.
The Event-to-Object graph (E2O) corresponding to the database event log $L_D$ can be defined as:
$$\textrm{E2O}(L_D) = (E \cup O, \textrm{EO})$$
Here, the nodes are the events ($E$) and the objects ($O$), and $EO$ (as retrieved from the log) is a subset of $E \times O$.
\label{def:e2ograph}
\end{mydef}
The E2O graph is obtained directly from the data without any transformation. The remaining steps in the construction
of an MVP model are the construction of the E2E multigraph and of the A2A multigraph.

\begin{mydef}[Sequence of related events]
Let $L_D = (E, O, C, A, \allowbreak \textrm{class}, \allowbreak \textrm{act}, \allowbreak \textrm{attr}, \allowbreak \textrm{EO}, \allowbreak \leq)$ be a database event log.
For $o \in O$, the following sequence of {\it related events} is defined:
$$
\widetilde{O}(o) = \{ e_1, \ldots, e_n \}
$$
such that
$\{ e_1, e_2, \ldots, e_n \} = \{ e ~ \arrowvert ~ (e, o) \in EO \}$ and $\forall_{1 \leq i < j \leq n} ~ ~ e_i < e_j$.
\end{mydef}

\begin{mydef}[E2E Multigraph]
Let $L_D = (E, O, C, A, \allowbreak \textrm{class}, \allowbreak \textrm{act}, \allowbreak \textrm{attr}, \allowbreak \textrm{EO}, \allowbreak \leq)$ be a database event log.
The Event-to-Event multigraph (E2E) on the database event log $L_D$ can be defined as:
$$E2E(L_D) = (E, F_E, \Pi^{E}_{\textrm{perf}})$$
Where the nodes are the events ($E$) and the set of edges $F_E$ is defined as:
$$F_E = \{ (e_1, e_2, o) \in E \times E \times O ~ \arrowvert ~ \exists_{2 \leq i \leq \arrowvert \widetilde{O}(o) \arrowvert} ~ \widetilde{O}_{i-1}(o) = e_1, \widetilde{O}_{i}(o) = e_2 \}$$
and $\Pi^{E}_{\textrm{perf}} : F_E \rightarrow \mathbb{R}^{+} \cup \{ 0 \}$, $\Pi^{E}_{\textrm{perf}}(e_1, e_2, o) = \textrm{attr}(e_2)(time) - \textrm{attr}(e_1)(time)$ associates each edge to a non-negative real number expressing its duration (performance).
\end{mydef}

The introduction of the set $F_E$ is useful for the definition of the A2A multigraph. Although it does not make sense to represent the overall E2E multigraph, involving relationships between all events, it may be useful
to display directly-follows relationships involving a small subgroup of events.

\begin{figure}[ht]
\centering
\includegraphics[width=250px]{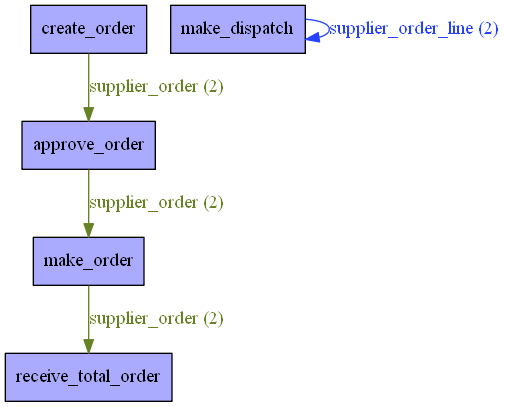}
\caption{Visualization of the A2A multigraph of an example database event log (erp.xoc test file). Activities are connected to activities; in the edge label, the (class) perspective along with the count of the occurrences has been reported.}
\end{figure}

\begin{mydef}[A2A Multigraph]
Let $L_D = (E, O, C, A, \allowbreak \textrm{class}, \allowbreak \textrm{act}, \allowbreak \textrm{attr}, \allowbreak \textrm{EO}, \allowbreak \leq)$ be a database event log.
Let $\textrm{AE} : A \times A \times C \rightarrow \mathcal{P}(E \times E \times O)$ such that for $a_1, a_2 \in A$ and $c \in C$:
$$\textrm{AE}(a_1, a_2, c) = \{ (e_1, e_2, o) \in F_E ~ \arrowvert ~ \textrm{act}(e_1) = a_1 \wedge \textrm{act}(e_2) = a_2 \wedge \textrm{class}(o) = c \}$$
The $\textrm{AE}$ function associates to each triple $(a_1, a_2, c)$ the set of all the events of the corresponding activities and classes.
The Activity-to-Activity multigraph (A2A) on the database event log $L_D$ can be defined as:
$$A2A(L_D) = (A, \allowbreak F_A, \allowbreak \Pi^{A}_{\textrm{count}}, \allowbreak \Pi^{A}_{\textrm{perf}})$$
Where the nodes are the activities ($A$), the set of edges $F_A$ is defined as:
$$F_A = \{ (a_1, a_2, c) \in A \times A \times C ~ \arrowvert ~ \textrm{AE}(a_1, a_2, c) \neq \emptyset \}$$
and:
\begin{itemize}
\item $\Pi^{A}_{\textrm{count}}(a_1, a_2, c) = \arrowvert \textrm{AE}(a_1, a_2, c) \arrowvert$ is the number of occurrences associated to the edge $(a_1, a_2, c) \in F_A$, that is the number
of corresponding edges contained in $\textrm{AE}(a_1, a_2, c)$.
\item $\Pi^{A}_{\textrm{perf}}(a_1, a_2, c) = \frac{\sum_{f_E \in \textrm{AE}(a_1, a_2, c)} \Pi^{E}_{\textrm{perf}}(f_E)}{\Pi^{A}_{\textrm{count}}(a_1, a_2, c)}$ is the performance associated to the edge $(a_1, a_2, c) \in F_A$,
that is the average of the duration of the corresponding edges contained in $\textrm{AE}(a_1, a_2, c)$. An high average duration may correspond to a bottleneck in the process.
\end{itemize}
\end{mydef}

The following definitions are useful for the representation of an MVP model, introducing clear start and end points for each class
and contributing to the possibility to filter out edges.

\begin{mydef}[Start and End Activities of a Class]
~ \\
Let $L_D = (E, O, C, A, \textrm{class}, \textrm{act}, \textrm{attr}, \allowbreak \textrm{EO}, \leq)$ be a database event log.
Let $c \in C$ be a class. The following functions are defined:
\begin{itemize}
\item $\textrm{START}_A(L_D) : C \rightarrow \mathcal{P}(A)$, $\textrm{START}_A(L_D)(c) = \{ \textrm{act}(\widetilde{O}_1(o)) ~ \arrowvert ~ o \in O \wedge \arrowvert \widetilde{O}(o) \arrowvert \geq 1 \wedge \textrm{class}(o) = c \}$ is the set of {\it start activities} of class $c$.
\item $\textrm{END}_A(L_D) : C \rightarrow \mathcal{P}(A)$, $\textrm{END}_A(L_D)(c) = \{ \textrm{act}(\widetilde{O}_{\arrowvert \widetilde{O}(o) \arrowvert}(o)) ~ \arrowvert ~ o \in O \wedge \arrowvert \widetilde{O}(o) \arrowvert \geq 1 \wedge \textrm{class}(o) = c \}$ is the set of {\it end activities} of class $c$.
\end{itemize}
\end{mydef}

\begin{mydef}[Dependency Threshold between Activities given a Class]
~ \\
Let $L_D = (E, O, C, A, \textrm{class}, \textrm{act}, \textrm{attr}, \allowbreak \textrm{EO}, \leq)$ be a database event log.
Let $F_A$ be the set of edges in $\textrm{A2A}(L_D)$.
For $(a_1,a_2,c) \in F_A$ it is possible to define a dependency measure in the following way:
$$\textrm{dep}_A(L_D) : F_A \rightarrow [0, 1]$$
$$\textrm{dep}_A(L_D)(a_1, a_2, c) =
\begin{cases}
\frac{\Pi^{A}_{\textrm{count}}(a_1, a_2, c)}{\Pi^{A}_{\textrm{count}}(a_1, a_2, c) + 1} & \textrm{if} ~ a_1 = a_2 \vee (a_2, a_1, c) \not\in F_A \\
\frac{\Pi^{A}_{\textrm{count}}(a_1, a_2, c) - \Pi^{A}_{\textrm{count}}(a_2, a_1, c)}{\Pi^{A}_{\textrm{count}}(a_1, a_2, c) + \Pi^{A}_{\textrm{count}}(a_2, a_1, c) + 1} & \textrm{if} ~ a_1 \neq a_2 \wedge (a_2, a_1, c) \in F_A \\
\end{cases}
$$
\end{mydef}

\begin{figure}[!t]
\centering
\includegraphics[width=200px]{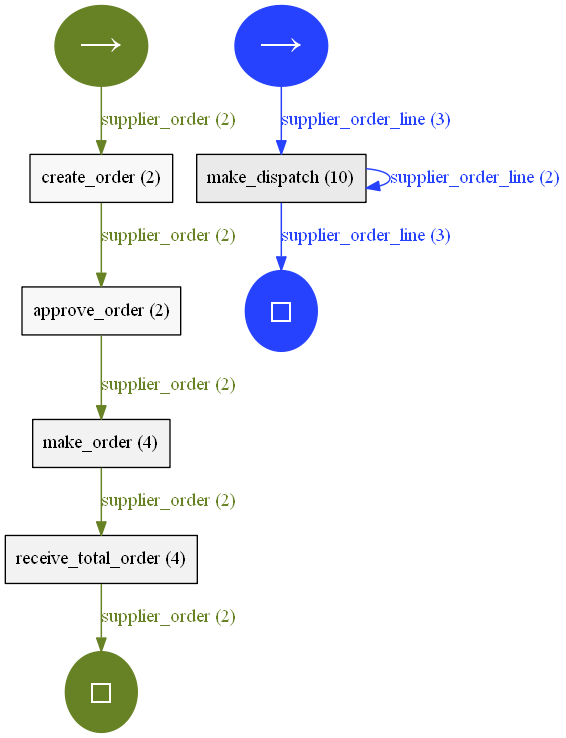}
\caption{Visualization of the MVP Model of an example database event log (erp.xoc test file). Activities are connected to activities; in the edge label, the (class) perspective along with the count of the occurrences has been reported. In addition to the A2A multigraph, start and end nodes are associated to each class; moreover, the edges that are reported in the A2A multigraph are filtered according to the dependency threshold (in this simple example, no arc is filtered).}
\end{figure}

\begin{mydef}[MVP Discovery]
Let $L_D = (E, O, C, A, \allowbreak \textrm{class}, \allowbreak \textrm{act}, \allowbreak  \textrm{attr}, \allowbreak \textrm{EO}, \allowbreak \leq)$ be a database event log.
We define as {\it MVP model} discovered from the log $L_D$, and we refer to it as $\textrm{MVP}(L_D)$, the following object:
\begin{align*}
\textrm{MVP}(L_D)=(L_D, \allowbreak \textrm{E2O}(L_D), \allowbreak \textrm{E2E}(L_D), \allowbreak \textrm{A2A}(L_D), \allowbreak \textrm{START}_A(L_D), \\
\textrm{END}_A(L_D), \allowbreak \textrm{dep}_A(L_D))
\end{align*}
\end{mydef}

Given a dependency threshold $d \in [-1, 1]$, a representation of an MVP model draws as many edges between a couple of activities $(a_1, a_2) \in A \times A$ as the number of classes $c \in C$ such that $dep_A(L_D)(a_1, a_2, c)$ is defined and
$\textrm{dep}_A(L_D)(a_1, a_2, c) \geq d$.

The visualization of a MVP model is valuable:
\begin{itemize}
\item An holistic view on the classes and the activities of the database, and on the order in which they happen, is provided.
\item Arcs are decorated with frequency/performance information. Frequency information helps to understand the most frequent paths of a process,
and performance information helps to discover the bottlenecks of a process.
\end{itemize}

\section{Viewpoints: Retrieval of DFGs and Logs}
\label{sec:projection}

The goal of this section is to provide some ways to retrieve, from an MVP model $\textrm{MVP}(L_D)$, built upon the database event log $L_D$,
a particular viewpoint on the model.

\begin{mydef}[Viewpoint]
\label{def:viewpoint}
~ \\
Let $L_D = (E, O, C, A, \allowbreak \textrm{class}, \allowbreak \textrm{act}, \allowbreak \textrm{attr}, \allowbreak \textrm{EO}, \leq)$ be a database event log.
Let
\begin{align*}
\textrm{MVP}(L_D)=(L_D, \allowbreak \textrm{E2O}(L_D), \allowbreak \textrm{E2E}(L_D), \allowbreak \textrm{A2A}(L_D), \allowbreak \textrm{START}_A(L_D), \\
\textrm{END}_A(L_D), \allowbreak \textrm{dep}_A(L_D))
\end{align*}
be an MVP model.
A viewpoint is a set of classes $V(L_D) \subseteq C$. A viewpoint is corresponding to a subset of edges in the E2E graph (that is $F_E$):
$$V_E(L_D) = \{ (e_1, e_2, o) \in F_E ~ \arrowvert ~ o \in V(L_D) \}$$
\end{mydef}

From a view, we can obtain two different final outputs.
\begin{itemize}
\item A Directly-Follows Graph (DFG), that includes the edges related to the classes contained in $V(L_D)$.
\item A classical event log, that includes all the events that are related to the objects having a class contained in $V(L_D)$.
\end{itemize}

\begin{figure}[ht]
\centering
\includegraphics[width=\textwidth]{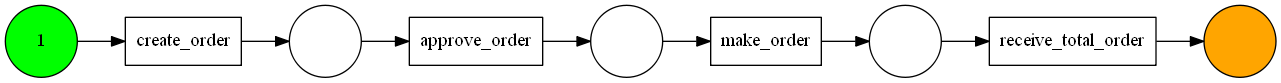}
\caption{Petri net extracted using Inductive Miner Directly-Follows from the projection of the MVP model extracted from the erp.xoc test file on the viewpoint containing the {\it supplier\_order} class.}
\label{fig:petrinet0}
\end{figure}

An example of projection could be found in Fig. \ref{fig:petrinet0},
where a singleton viewpoint containing a single class is chosen, a Directly-Follows Graph is obtained and a Petri net is obtained through the application of Inductive Miner Directly-Follows.

\subsection{Projection on a Directly-Follows Graph}

The concept of a Directly-Follows Graph is introduced in the following definition:

\begin{mydef}[DFG]
A Directly-Follows Graph is a weighted directed graph:
$$\textrm{DFG} = (N_{\textrm{DFG}}, E_{\textrm{DFG}}, c_{\textrm{DFG}})$$
Where $N_{\textrm{DFG}}$ (the nodes) are the activities, and $E_{\textrm{DFG}} \subseteq N_{\textrm{DFG}} \times N_{\textrm{DFG}}$ is the set of all the edges between activities that happened in direct succession,
and $c_{\textrm{DFG}} : E_{\textrm{DFG}} \rightarrow \mathbb{R}^{+}$ is the count function that aims to represent how many times two different activities happened in direct succession.
\end{mydef}

From a viewpoint $V(L_D)$ on $\textrm{MVP}(L_D)$, it is possible to obtain a Directly-Follows Graph as explained in the following definition:

\begin{mydef}[DFG given a viewpoint]
Given a viewpoint $V(L_D)$ on $\textrm{MVP}(L_D)$, a DFG could be obtained taking:
\begin{itemize}
\item $N_{\textrm{DFG}} = \bigcup_{(e_1, e_2, o) \in V_E(L_D)} \{ \textrm{act}(e_1), \textrm{act}(e_2) \}$
\item $E_{\textrm{DFG}} = \bigcup_{(e_1, e_2, o) \in V_E(L_D)} \{ (\textrm{act}(e_1), \textrm{act}(e_2)) \}$
\item $c_{\textrm{DFG}}(a_1, a_2) = \arrowvert \{ (e_1, e_2, o) \in V_E(L_D) ~ \arrowvert ~ \textrm{act}(e_1) = a_1 \wedge \textrm{act}(e_2) = a_2 \} \arrowvert$
\end{itemize}
where $V_E(L_D)$ is obtained as in Def. \ref{def:viewpoint}.
\end{mydef}

An example could be provided. Let $V_E(L_D) = \{ (e_1, e_2, o_1), \allowbreak (e_2, e_3, o_1), \allowbreak (e_4, e_5, o_2), \allowbreak (e_5, e_6, o_2) \}$ be the set of edges associated
to a viewpoint, such that:
\begin{itemize}
\item $\textrm{act}(e_1) = \textrm{act}(e_4) = A$
\item $\textrm{act}(e_2) = \textrm{act}(e_5) = B$
\item $\textrm{act}(e_3) = C$
\item $\textrm{act}(e_6) = D$
\end{itemize}
Then the DFG is such that $N_{\textrm{DFG}} = \{ A, B, C, D \}$, $E_{\textrm{DFG}} = \{ (A, B), (B, C), (B, D) \}$,
$c_{\textrm{DFG}}(B, C) = c_{\textrm{DFG}}(B, D) = 1$, $c_{\textrm{DFG}}(A, B) = 2$.

\subsection{Projection on a Log}

The projection of an MVP model $\textrm{MVP}(L_D)$ obtained from a database event log $L_D$ to a classical event log (see Section \ref{sec:background})
could be introduced when a viewpoint $V(L_D) \subseteq C$ is chosen.

Indeed, the information contained in an MVP model could be used to determine a case notion $C_D$ that is used to transform the database event log $L_D$ into a classical event log,
in a way that events belonging to the same process execution can be grouped.

The definition of case notion on database event logs could be introduced:
\begin{mydef}[Case Notion]
Let $L_D = (E, O, C, A, \textrm{class}, \allowbreak \textrm{act}, \allowbreak \textrm{attr}, \allowbreak \textrm{EO}, \allowbreak \leq)$ be a database event log.
A {\it case notion} is a set of sets of events $C_D \subseteq \mathcal{P}(E) \setminus \{ \emptyset \}$.
\end{mydef}
The case notion does not need to cover all the events contained in $E$, moreover
the intersection between sets of events contained in the case notion may also not be empty.

An example of case notion could be provided. Let $E = \{ e_1, e_2, e_3, e_4, e_5, e_6 \}$ be a set of events. Then a case notion might be
$C_D = \{ \{ e_1, e_2 \}, \{ e_3, e_4 \}, \{ e_1, e_4, e_5 \} \}$. Let's note that the union of all these sets is not $E$, and the intersection
between $\{ e_1, e_2 \}$ and $\{ e_1, e_4, e_5 \}$ is not empty. 

When a case notion is defined, it is possible to define the projection function from the database event log $L_D$ to the classical event log.
An assumption is that the case notion $C_D$ is contained in the universe of case identifiers $\mathcal{U}_C$. This helps to define the function
$\textrm{case\_ev}$ in a simpler way.

\begin{mydef}[Projection function]
Let $L_D = (E, O, C, A, \textrm{class}, \textrm{act}, \textrm{attr}, \allowbreak \textrm{EO}, \leq)$ be an event log in a database context.
Let $C_D \subseteq \mathcal{P}(E) \setminus \{ \emptyset \}$ be a case notion. Then it is possible to define a projection function
from a database event log to a classical event log as:
$$\textrm{proj}(L_D,C_D) = (C_D,E,A,\textrm{case\_ev},\textrm{act},\textrm{attr},\leq)$$
where $\textrm{case\_ev} \in C_D \rightarrow \mathcal{P}(E) \setminus \{ \emptyset \}$ is such that for all $c \in C_D$, $\textrm{case\_ev}(c) = c$.
\end{mydef}

Given an MVP model $\textrm{MVP}(L_D)$ and a viewpoint $V(L_D)$ on that defines a set of edges $V_E(L_D)$ in the E2E multigraph, a case notion is defined as:
$$\textrm{C}_D = \left \{ \bigcup_{o' \in O, \textrm{class}(o') \in V(L_D), \widetilde{O}(o) \cap \widetilde{O}(o') \neq \emptyset } \widetilde{O}(o') ~ ~ \arrowvert ~ o \in O, \textrm{class}(o) \in V(L_D) \right \}$$
A classical event log is obtained as $L = \textrm{proj}(L_D, C_D)$.

\section{Tool}
\label{sec:tool}

MVP Models have been implemented in a feature branch of the PM4Py Process Mining library\footnote{The repository can be accessed at the URL \url{https://github.com/Javert899/pm4py-source}} \cite{berti2019process}.
The architecture of the tool provides a clear separation between the management of the log (object), the MVP discovery algorithm and the MVP visualization. Moreover,
utilities have been provided to generate a database event log and to visualize the E2O and the E2E multigraphs.  A reference technical manual with a description of the features provided in
tool is contained in \url{http://www.alessandroberti.it/technical.pdf}.

%

\begin{figure}[ht]
\centering
\tiny
\begin{tabular}{|l|l|l|l|l|}
\toprule
                   event\_id &         event\_activity &     event\_timestamp &    supplier\_order &    supplier\_order\_line \\
\midrule
             create\_order15 &           create\_order & 2016-10-21 11:38:26 &  supplier\_order15 &                    NaN \\
            approve\_order15 &          approve\_order & 2016-10-21 11:38:53 &  supplier\_order15 &                    NaN \\
             make\_order1027 &             make\_order & 2016-10-21 11:40:00 &               NaN &                    NaN \\
             make\_order1027 &             make\_order & 2016-10-21 11:40:00 &  supplier\_order15 &                    NaN \\
            make\_dispatch19 &          make\_dispatch & 2016-10-21 11:42:31 &               NaN &                    NaN \\
            make\_dispatch19 &          make\_dispatch & 2016-10-21 11:42:31 &               NaN &  supplier\_order\_line22 \\
            make\_dispatch18 &          make\_dispatch & 2016-10-21 11:42:31 &               NaN &                    NaN \\
            make\_dispatch18 &          make\_dispatch & 2016-10-21 11:42:31 &               NaN &  supplier\_order\_line23 \\
  receive\_partial\_order1028 &  receive\_partial\_order & 2016-10-21 11:43:00 &               NaN &                    NaN \\
            make\_dispatch21 &          make\_dispatch & 2016-10-21 11:44:25 &               NaN &                    NaN \\
            make\_dispatch21 &          make\_dispatch & 2016-10-21 11:44:25 &               NaN &  supplier\_order\_line22 \\
            make\_dispatch20 &          make\_dispatch & 2016-10-21 11:44:25 &               NaN &                    NaN \\
            make\_dispatch20 &          make\_dispatch & 2016-10-21 11:44:25 &               NaN &  supplier\_order\_line23 \\
    receive\_total\_order1029 &    receive\_total\_order & 2016-10-21 11:45:00 &               NaN &                    NaN \\
    receive\_total\_order1029 &    receive\_total\_order & 2016-10-21 11:45:00 &  supplier\_order15 &                    NaN \\
           create\_invoice17 &         create\_invoice & 2016-10-21 11:45:46 &               NaN &                    NaN \\
           create\_invoice17 &         create\_invoice & 2016-10-21 11:45:46 &               NaN &                    NaN \\
           create\_payment11 &         create\_payment & 2016-10-21 11:46:14 &               NaN &                    NaN \\
           create\_payment11 &         create\_payment & 2016-10-21 11:46:14 &               NaN &                    NaN \\
           create\_payment12 &         create\_payment & 2016-10-21 11:46:29 &               NaN &                    NaN \\
           create\_payment12 &         create\_payment & 2016-10-21 11:46:29 &               NaN &                    NaN \\
             create\_order16 &           create\_order & 2016-10-21 11:56:35 &  supplier\_order16 &                    NaN \\
            approve\_order16 &          approve\_order & 2016-10-21 11:56:50 &  supplier\_order16 &                    NaN \\
             make\_order1033 &             make\_order & 2016-10-21 11:57:00 &               NaN &                    NaN \\
             make\_order1033 &             make\_order & 2016-10-21 11:57:00 &  supplier\_order16 &                    NaN \\
            make\_dispatch22 &          make\_dispatch & 2016-10-21 11:57:28 &               NaN &                    NaN \\
            make\_dispatch22 &          make\_dispatch & 2016-10-21 11:57:28 &               NaN &  supplier\_order\_line24 \\
    receive\_total\_order1034 &    receive\_total\_order & 2016-10-21 11:58:00 &               NaN &                    NaN \\
    receive\_total\_order1034 &    receive\_total\_order & 2016-10-21 11:58:00 &  supplier\_order16 &                    NaN \\
           create\_invoice18 &         create\_invoice & 2016-10-21 11:58:36 &               NaN &                    NaN \\
           create\_invoice18 &         create\_invoice & 2016-10-21 11:58:36 &               NaN &                    NaN \\
\bottomrule
\end{tabular}
\caption{Representation of a database log in the Parquet columnar format.
This is different from classic CSV event logs since the same event is repeated in multiple rows, one for each related object.
Both information related to events and related objects are columns of a table.
This structure reflects the way information is stored inside the in-memory Pandas dataframe that supports the filtering and the discovery operations.}
\label{fig:tabularParquet}
\end{figure}

The provided features are:
\begin{itemize}
\item {\bf Log management}: log importing (XOC, OpenSLEX, Parquet), log exporting (XOC, Parquet).
\item {\bf Model discovery}: discovery of a MVP model with frequency or performance decoration.
\item {\bf Visualization}: E2O graphs, E2E and A2A multigraphs, MVP models (with possibility to filter out edges with a dependency measure that is below the threshold).
\item {\bf Database log generation}: the option to generate an MVP model specifying the number of events,
activities, classes and a number of objects for each class is provided.
\item {\bf Projection on a Viewpoint}: projection on a DFG and on a log.
\item {\bf Storage of MVP models}: importing/exporting of MVP models into a dump file.
\end{itemize}

Example logs are provided in the tests folder of the repository, in particular:
\begin{itemize}
\item {\it logOpportunities.parquet} is the database log used in the assessment and extracted from the Dynamics CRM system.
\item {\it metamodel.slexmm} provides the OpenSLEX metamodel of the Concert database.
\item {\it erp.xoc} provides an example XOC log extracted from a Dollibar ERP system.
\end{itemize}

The storage used in the tool for the database event logs is represented in Fig. \ref{fig:tabularParquet}.
Several rows are associated to an event, and contain an ID, an activity, a timestamp, and a single object identifier in the column corresponding to its class.
This permits to store events in a tabular format using basic types for the columns (strings, integers, dates) to maximize the query performance.

\pgfplotstableread[col sep=space,row sep=newline,header=true]{
x   y
30 0.56
58 0.78
112 1.16
249 2.05
}\speedMVP

\pgfplotstableread[col sep=space,row sep=newline,header=true]{
x   y
30 12.0
58 25.0
112 73.0
249 260.0
}\speedOCBC

\pgfplotstableread[col sep=space,row sep=newline,header=true]{
x   y
155 11
326 18
661 30
}\sizeMVP

\pgfplotstableread[col sep=space,row sep=newline,header=true]{
x   y
155 29413
326 130760
661 524305
}\sizeOCBC

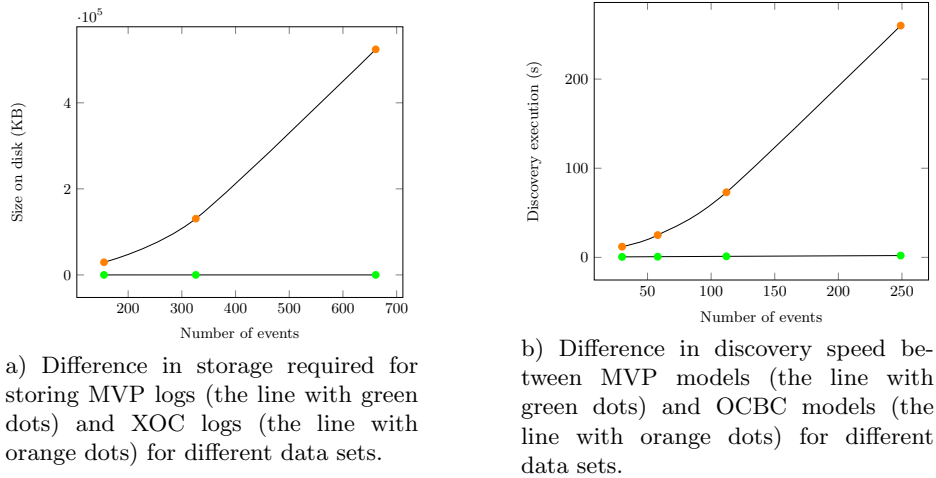
\begin{figure}
    \begin{minipage}{0.45\textwidth}
\resizebox{\columnwidth}{!}{%
\begin{tikzpicture}
  \begin{axis}[xlabel={Number of events}, ylabel={Size on disk (KB)}]
\addplot[smooth] table {\sizeMVP};
\addplot[smooth] table {\sizeOCBC};

\addplot[only marks,mark=*,mark options={color=green}] table {\sizeMVP};
\addplot[only marks,mark=*,mark options={color=orange}] table {\sizeOCBC};
\end{axis}
\end{tikzpicture}
} \\
a) Difference in storage required for storing MVP logs (the line with green dots) and XOC logs (the line with orange dots) for different data sets.

    \end{minipage}
	\begin{minipage}{0.1\textwidth}
	~
	\end{minipage}
    \begin{minipage}{0.45\textwidth}
\resizebox{\columnwidth}{!}{%
\begin{tikzpicture}
  \begin{axis}[xlabel={Number of events}, ylabel={Discovery execution (s)}]
\addplot[smooth] table {\speedMVP};
\addplot[smooth] table {\speedOCBC};

\addplot[only marks,mark=*,mark options={color=green}] table {\speedMVP};
\addplot[only marks,mark=*,mark options={color=orange}] table {\speedOCBC};
\end{axis}
\end{tikzpicture}
}
b) Difference in discovery speed between MVP models (the line with green dots) and OCBC models (the line with orange dots) for different data sets.
\end{minipage}
\caption{Comparison between MVP models and OCBC models, in (a) size on disk of the proposed log storage b) discovery speed. The XOC format that is tested is the one described in \cite{van2017object}.}
\label{fig:comparisonMVPOCBC}
\end{figure}

\section{Assessment}
\label{sec:assessment}

The assessment of MVP models will show how they perform against two competing approaches (OCBC models \cite{li2017automatic} and OpenSLEX \cite{de2018connecting}),
both considering the execution time and the simplicity/readability/amount of information contained in the models.
These approaches have been chosen because are very recent.
The scalability of the MVP models implementation contained in PM4Py will be analyzed on some synthetic logs.
Moreover, an assessment using a real information system
(Microsoft Dynamics CRM) will be shown.

\begin{figure}[ht]
\centering
\includegraphics[width=\textwidth]{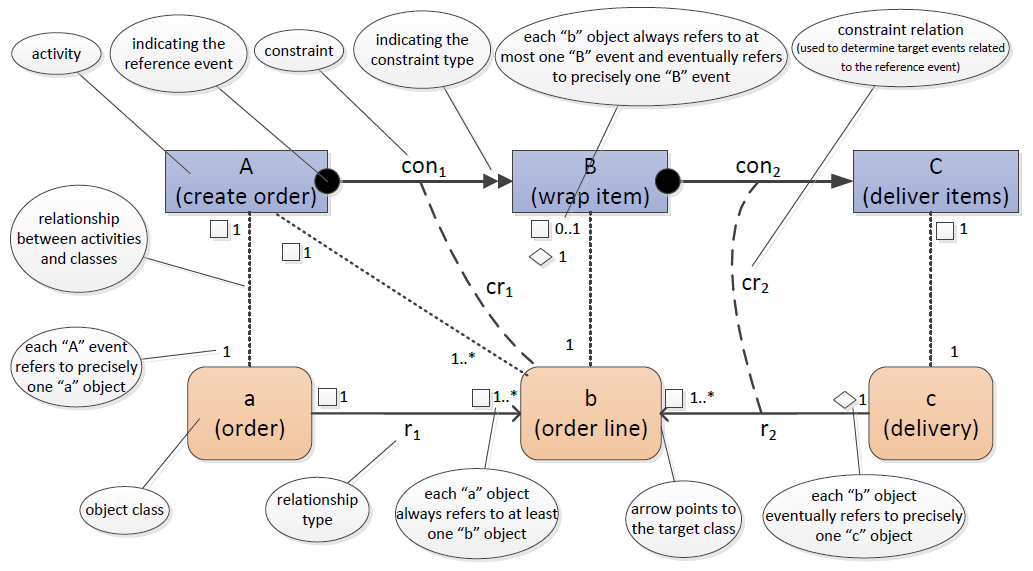}
\caption{Usability assessment: example model illustrating the main ingredients of OCBC models.}
\label{fig:explanationOCBC}
\end{figure}

\begin{figure}[ht]
\centering
\includegraphics[width=\textwidth]{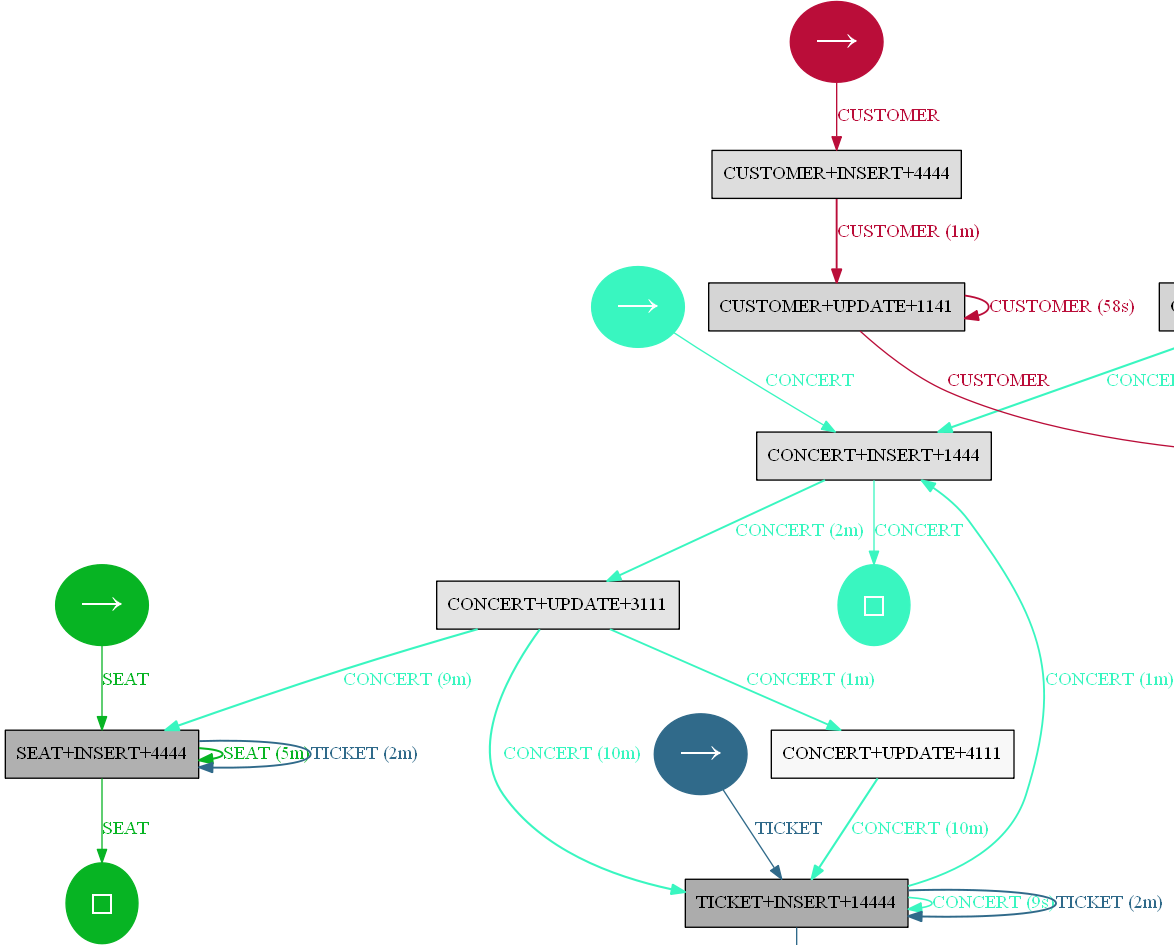}
\caption{Usability assessment: performance/bottleneck visualization provided by MVP models.}
\label{fig:explanationMVC}
\end{figure}

\subsection{Comparison with Related Approaches}

OCBC models are powerful descriptions of the relationships between activities and classes at the database level.
Having said that, OCBC models have scalability issues:
XOC logs as proposed in \cite{van2017object} store a snapshot of the object model per event,
this becomes quickly unfeasible also for few hundred database events
(as it will shown in the assessment).
In \cite{de2018extracting}, an updated version of XOC, storing in each event the updates to the object model, has been proposed,
although a XOC event log in such format is not public available.
Moreover, the final visualization (even if it can be filtered on some types of constraints) lacks understandability.
A serious issue is the lack of support for frequency/performance decoration (i.e. the number of occurrences of the arc, the time passed between activity).

OpenSLEX provides a way to ingest a database event log into a meta-model instance that is easier to query.
An issue is that they do not offer any visual clue on underlying relationships between activities, making the life difficult to the user in first instance.
Moreover, although the queries on OpenSLEX are easier than the queries done directly on the database, there is an effort required by the user to understand the concepts described
in \cite{de2018connecting}.

\subsection{Evaluation of the Execution Time}

OCBC models (tests have been performed on the original version of XOC, proposed in \cite{van2017object}, for which several logs are public available)
have scalability problems with regards to the log format (XOC), that requires the storage, for each event, of the status of the entire object model.
In Fig. \ref{fig:comparisonMVPOCBC} (a), the size on disk (in KB) of a log containing the specified number of events has been considered.
To obtain an OCBC model, for a log with just $661$ events, a XOC log of $512$ MB is required, that is $17476$ greater than the amount of disk space that permits to discover an MVP model from the same database.

In Fig. \ref{fig:comparisonMVPOCBC} (b), the execution speed of the discovery procedure has been compared between MVP models and OCBC models.
To obtain an OCBC model, for a log with just $249$ events, a time of $4$ minutes and $20$ seconds is required, while for discovering an MVP model from the same database
$2$ seconds are required, that is $127$ times faster.
This result holds also for the new version of the XOC format \cite{de2018extracting}, since in the later version the object model is built in-memory
by the discovery algorithm starting from the updates described in the event log.

With OpenSLEX, the comparison is more tight, although the final goal is different: OpenSLEX require a query to get a classical event log,
while MVP models do not require this effort. There is only a meta-model instance available in public, that has been extracted from a synthetic database on concert management.
The amount of storage required to store the OpenSLEX instance is $11$ MB, while the amount of storage that permits to discover an MVP model is $222$ KB (storing using Parquet format).
To compare the (time) performance of OpenSLEX and MVP models, a query on the OpenSLEX instance needs to be performed.
Starting in both cases from the concert database instance of the OpenSLEX meta-model,
if an example query\footnote{The example query is provided in the technical report available at the address \url{http://www.alessandroberti.it/technical.pdf}} is chosen, the execution speed of the query on the OpenSLEX instance is $1.73$ seconds.
The time needed for
MVP models to obtain a complete model with frequency and performance information is $1.96$ seconds, this without requiring the specification of any query by the user.

\pgfplotstableread[col sep=space,row sep=newline,header=true]{
x   y
20.0 3.34
40.0 3.45
60.0 6.27
80.0 6.94
120.0 14.11
140.0 23.35
}\changingNActivities

\pgfplotstableread[col sep=space,row sep=newline,header=true]{
x   y
2000.0 3.62
4000.0 3.61
6000.0 3.76
8000.0 3.67
10000.0 3.9
12000.0 4.15
14000.0 4.06
}\changingNObjectsPerClass

\pgfplotstableread[col sep=space,row sep=newline,header=true]{
x   y
5.0 0.39
10.0 0.93
15.0 1.5
20.0 2.2
25.0 2.97
30.0 3.74
35.0 4.49
}\changingNClasses

\pgfplotstableread[col sep=space,row sep=newline,header=true]{
x   y
50000.0 2.15
100000.0 3.61
150000.0 5.18
200000.0 6.82
250000.0 8.43
300000.0 10.04
}\changingNEvents

\begin{figure}
    \begin{minipage}[t]{0.45\textwidth}
\resizebox{\columnwidth}{!}{%
\begin{tikzpicture}
  \begin{axis}[xlabel={Number of activities}, ylabel={Discovery time (s)}]
\addplot[smooth] table {\changingNActivities};
\addplot[only marks,mark=*,mark options={color=blue}] table {\changingNActivities};
\end{axis}
\end{tikzpicture}
} \\
a) Performance of MVP discovery with the increase of the number of activities, when the number of classes, the number of objects per class and the number of events in the log is kept fixed: n\_events = 100000, n\_classes = 10 and n\_objects\_per\_class = 1000. The logs have been obtained through the generator included in PM4Py. The execution time (in seconds) grows quadratically with the number of activities.
    \end{minipage}
	\begin{minipage}[t]{0.1\textwidth}
	~
	\end{minipage}
    \begin{minipage}[t]{0.45\textwidth}
\resizebox{\columnwidth}{!}{%
\begin{tikzpicture}
  \begin{axis}[xlabel={Number of objects per class}, ylabel={Discovery time (s)}]
\addplot[smooth] table {\changingNObjectsPerClass};
\addplot[only marks,mark=*,mark options={color=blue}] table {\changingNObjectsPerClass};
\end{axis}
\end{tikzpicture}
} \\
b) Performance of MVP discovery with the increase of the number of objects per class, when the number of classes, the number of activities and the number of events in the log is kept fixed: n\_events = 100000, n\_classes = 10 and n\_activities = 40. The logs have been obtained through the generator included in PM4Py.
\end{minipage} \\
    \begin{minipage}[t]{0.45\textwidth}
\resizebox{\columnwidth}{!}{%
\begin{tikzpicture}
  \begin{axis}[xlabel={Number of classes}, ylabel={Discovery time (s)}]
\addplot[smooth] table {\changingNClasses};
\addplot[only marks,mark=*,mark options={color=blue}] table {\changingNClasses};
\end{axis}
\end{tikzpicture}
} \\
c) Performance of MVP discovery with the increase of the number of classes, when the number of objects per class, the number of activities and the number of events in the log is kept fixed: n\_events = 10000, n\_activities = 40 and n\_objects\_per\_class = 6000. The logs have been obtained through the generator included in PM4Py. The execution time (in seconds) grows linearly with the number of classes.
    \end{minipage}
	\begin{minipage}[t]{0.1\textwidth}
	~
	\end{minipage}
    \begin{minipage}[t]{0.45\textwidth}
\resizebox{\columnwidth}{!}{%
\begin{tikzpicture}
  \begin{axis}[xlabel={Number of events}, ylabel={Discovery time (s)}]
\addplot[smooth] table {\changingNEvents};
\addplot[only marks,mark=*,mark options={color=blue}] table {\changingNEvents};
\end{axis}
\end{tikzpicture}
} \\
d) Performance of MVP discovery with the increase of the number of events in the log, when the number of classes, the number of objects per class and the number of activities in the log is kept fixed: n\_activities = 40, n\_classes = 10 and n\_objects\_per\_class = 6000. The logs have been obtained through the generator included in PM4Py. The execution time (in seconds) grows linearly with the number of events.
\end{minipage}
\caption{Scalability assessment of MVP models, with regards to a) the number of activities b) the number of objects per class c) the number of classes d) the number of events in the log.}
\label{fig:multipleAssessmentScalability}
\end{figure}

\subsection{Usability of the Approaches}
\label{sec:usability}

OCBC and MVP models both provide ways to discover a model on top of database event logs. The amount of information and constraints extracted by OCBC models is
very high, and although filtering is provided to keep only some constraints, the resulting process model is complex to understand.
An explanation of some ingredients of OCBC models is provided by \cite{li2017automatic} and represented in Fig. \ref{fig:explanationOCBC}.
This amount of information is insane, but it is very difficult also for a process analyst to be able to understand it without proper training.

Moreover, this class of models does not provide frequency/performance information, that can be useful for the process analyst in order to detect the bottlenecks.
MVP models, as represented in Fig. \ref{fig:explanationMVC}, can provide a graph in which edges are decorated by frequency/performance information.
With MVP models, the following features are also available, that are not available on OCBC models:
\begin{itemize}
\item Projection to a classic directly-follows graph to be used with techniques like Inductive Miner Directly-Follows and the Heuristics Miner.
\item Projection to a classical event log to be used with mainstream process mining techniques.
\end{itemize}

OpenSLEX do not provide a visualization of a process model on top of the database, but require to the user the specification of a query to retrieve
an event log. This is an unavoidable step, and requires time and expertise by the user. So, the retrieval of a process model using MVP requires less time
and less knowledge than the retrieval from OpenSLEX.

\subsection{Scalability of the Approach}
\label{sec:selfAssessment}

In the previous section, MVP models have shown greater scalability and usability in comparison to some competing approaches.
In this section, the goal is to understand more clearly the performance of the current implementation.

An assessment of the approach on simulated logs (through the log generator included in PM4Py) has been done (see Fig. \ref{fig:multipleAssessmentScalability}) to see which variables influence the execution time in a quadratic way,
and which variables influence it in a linear way:
\begin{itemize}
\item a) assesses the performance of MVP discovery with the increase of the number of activities, when the number of classes, the number of objects per class and the number of events in the log is kept fixed. The execution time (in seconds) grows quadratically with the number of activities.
\item b) assesses the performance of MVP discovery with the increase of the number of objects per class, when the number of classes, the number of activities and the number of events in the log is kept fixed. Although the behavior in the figure looks erratic, the execution time (in seconds) grows asymptotically linearly with the number of objects per class.
\item c) assesses the performance of MVP discovery with the increase of the number of classes, when the number of objects per class, the number of activities and the number of events in the log is kept fixed. The execution time (in seconds) grows linearly with the number of classes.
\item d) assesses the performance of MVP discovery with the increase of the number of events in the log, when the number of classes, the number of objects per class and the number of activities in the log is kept fixed. The execution time (in seconds) grows linearly with the number of events.
\end{itemize}

\begin{figure}
\centering
\includegraphics[width=\textwidth]{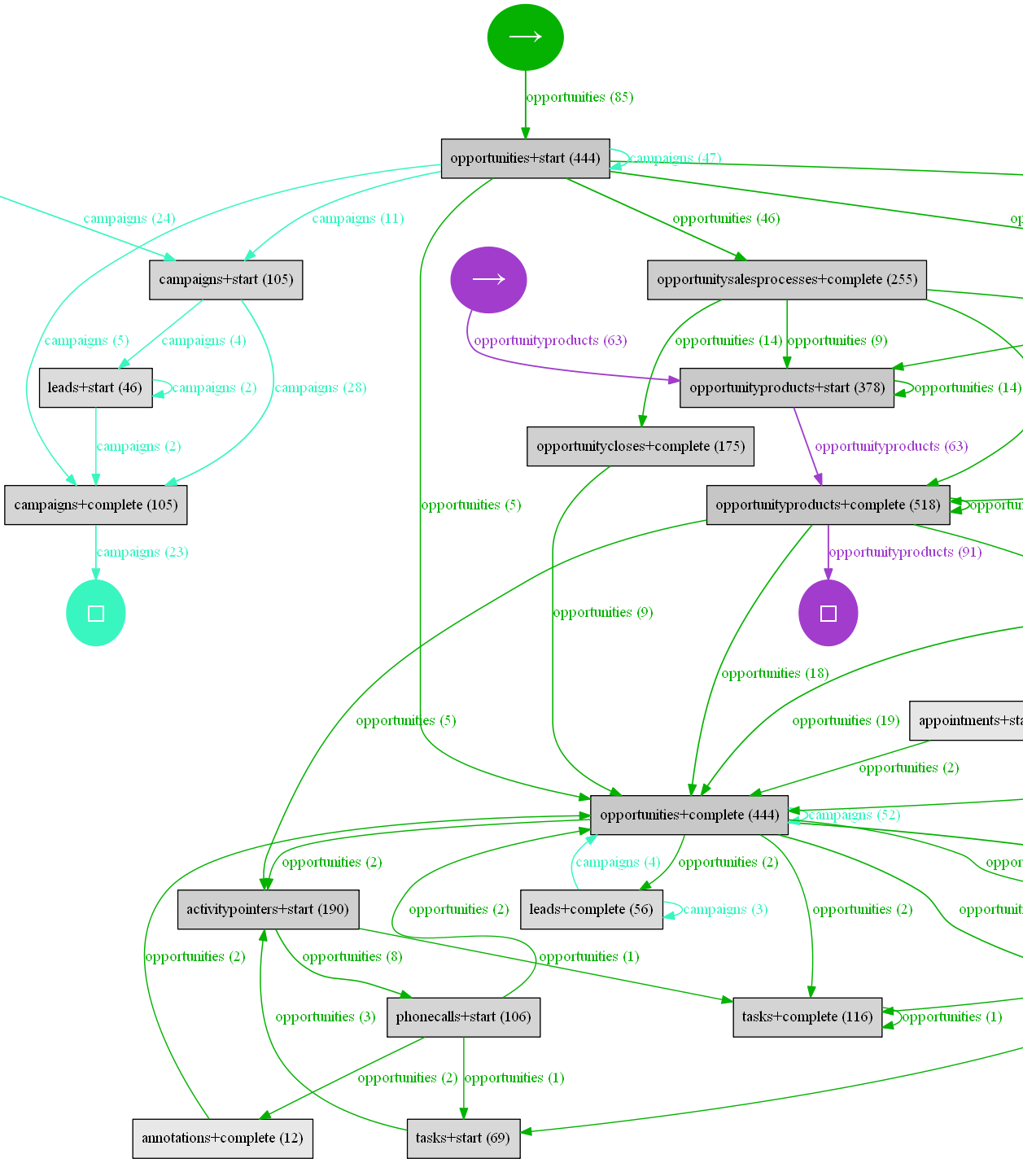}
\caption{Representation in one model of four perspectives (opportunities, opportunityproducts, campaigns, connections) of the Dynamics CRM database (only part of the diagram is reported).}
\label{fig:fourPerspectives}
\end{figure}

\begin{figure}[ht]
\centering
\includegraphics[width=\textwidth]{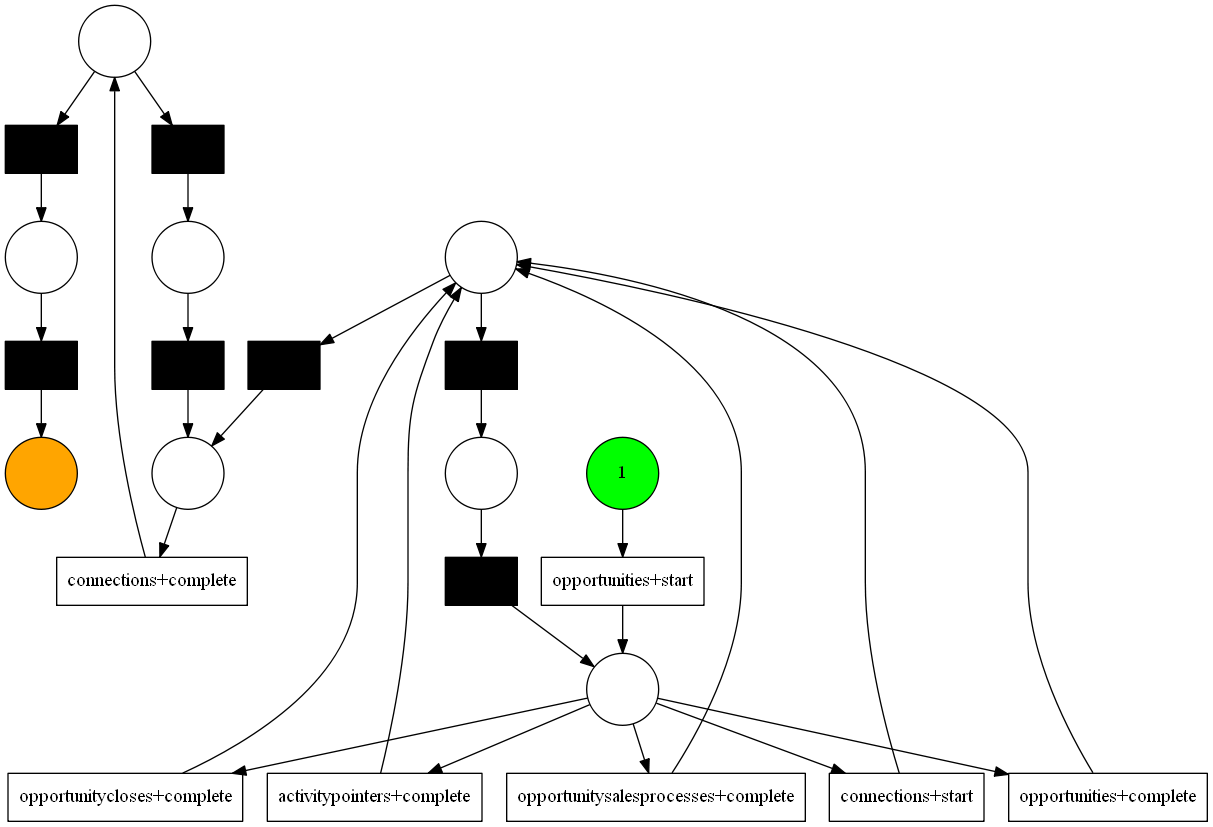}
\caption{Petri net obtained by applying Inductive Miner Directly-Follows on the projection of the Dynamics CRM MVP model on a viewpoint containing only the 'opportunities' class.
Black boxes represent invisible transitions (transitions that can be executed without correspondence with the activities of the log).
}
\label{fig:opportunitiesPetri}
\end{figure}

\subsection{Evaluation using CRM Data}
\label{sec:assessmentReal}
This section presents a study of data extracted from a Microsoft Dynamics CRM demo and analyzed using MVP discovery. A Customer Relationship Management system (CRM) \cite{buttle2004customer} is an information
system used to manage the commercial lifecycle of an organization, including management of customers, opportunities and marketing campaigns.

Many companies actually involve a CRM system for helping business and sales people to coordinate, share information and goals.
Data extracted from Microsoft Dynamics CRM is particularly interesting since this product manages several processes of the business side, providing the possibility to define workflows
and to measure KPI also through connection to the Microsoft Power BI business intelligence tool.
For evaluation purposes, a database log has been generated containing data extracted from a Dynamics CRM demo.
The database supporting the system contains several entities, and each entity contains several entries related to activities happening in the CRM.
Each entry could be described by a unique identifier (UUID), the timestamp of creation/change, the resource that created/modified the entry, and some UUIDs of other
entries belonging to the same or to different entities. Moreover, each entry is uniquely associated with the entity it belongs to.

The following strategy has been pursued in order to generate a log:
\begin{itemize}
\item For each entry belonging to an entity, two events have been associated: creation event (with the timestamp of creation and lifecycle {\it start}) and modify event
(with the timestamp of modification and lifecycle {\it complete}).
\item Each entry belonging to an entity has also been associated with an object.
\item Relationships between events and objects are created accordingly to the relationships expressed by the entries (an entry may cite several UUIDs of other entries stored in the database).
\end{itemize}
The previous construction means that for the same entry there are two events (start+complete) and one object in the log.

The database log contains 5863 events, 4413 objects, 120 activities and 80 object classes, and could be stored in a 386 KB Parquet file.
The complete MVP model (525 edges) can be calculated and represented in 5 seconds.
Taking a viewpoint containing only classes related to opportunities management (e.g. {\it opportunities}, {\it opportunityproducts}, {\it campaigns}, {\it connections}), and choosing the frequency metric, the model
obtained is represented in Fig. \ref{fig:fourPerspectives}. Projecting the MVP model on a viewpoint containing only the 'opportunities' class and applying Inductive Miner Directly-Follows,
the process model represented in Fig. \ref{fig:opportunitiesPetri} is obtained.

Being able to handle this complex database schema and visualize a model that unifies the different classes, in a very reasonable time, is a thing that is impossible
with competing approaches like OCBC models (due to severe scalability issues) and OpenSLEX instances (because the resulting query would be more complex than selecting a viewpoint from an holistic model).

\section{Conclusion}

This paper introduces Multiple Viewpoint models (MVP), providing an holistic view on a process supported by a database.
MVP models are annotated with frequency and performance measures (e.g., delays), supporting the detection of the most frequent paths and of the bottlenecks
without the specification of any case notion.
At the same time, a viewpoint (a non-empty subset of classes) can be chosen on the MVP models in order to get classical objects (DFGs and logs) to use with the mainstream process mining techniques.
Hence, the holistic view of the MVP model is complemented by detailed viewpoint models using conventional notations like Petri nets, BPMN models and process trees.
This possibility is not provided by the competing techniques described in Section \ref{sec:rw}: the focus is either on the specification of a case notion, or on the retrieval
of an artifact-centric model.
However, the relationships between the classes are not calculated and represented in the process model, in contrast to the technique of OCBC models. Moreover, the lack of a clear
execution semantic of MVP models makes the application of conformance checking techniques possible only after choosing a viewpoint, in contrast with the OCBC technique that provides
a (theoretically) powerful conformance checking approach on top of the model.

An MVP model has been
discovered from a database event log extracted from a Microsoft Dynamics Customer Relationship Management (CRM) system, showing the possibility to apply the
techniques described in this paper to real-life information systems.

Moreover, a comparison considering execution time and usability has been performed
against two of the most recent process mining approaches on databases (OpenSLEX and OCBC models); MVP models are relatively well performing and easy to use,
since an holistic view could be obtained without any effort from the user, and any viewpoint could be chosen on the MVP model.

Scalability testing proved that MVP models scale linearly with the number of events, classes and asymptotically linearly with the number of objects contained in the database event log, while they scale
quadratically with the number of activities (due to the edges calculation in the A2A multigraph). This is a remarkable result in comparison to OCBC models that show
an exponential complexity on the number of events (they become unmanageable also for a small number of events).

The techniques described in this paper could in principle be implemented starting from the logs of any relational database.
Hence MVP discovery supports process mining analysis directly from real-life complex information systems.

%
%
%
%
\bibliographystyle{splncs04}
\bibliography{starstar}

\end{document}